\documentstyle[preprint,aps]{revtex}
\draft
\preprint{Phys. Rev. C (Aug., 1996) in press.}
\begin{document}
 
\title{$p_t$ dependence of transverse flow in relativistic
heavy-ion collisions}
\bigskip
\author{Bao-An Li, C.M. Ko and G.Q. Li}
\address{Cyclotron Institute and Physics Department,\\ 
Texas A\&M University, College Station, TX 77843, USA}
\maketitle
 
\begin{quote}
The strength of transverse flow is examined as
a function of transverse momentum $p_t$ using a simple, transversely 
moving thermal model and a more realistic, relativistic transport model (ART).
It is shown that the $p_t$ dependence reveals useful information about the 
collective flow that is complementary to that obtained from the 
standard in-plane transverse momentum analysis. Interesting features of 
using the $p_t$ dependence to study the equation of state of the superdense 
hadronic matter formed in relativistic heavy-ion collisions are demonstrated. 
\end{quote}
\newpage
Compressional shock waves created in relativistic heavy-ion collisions
were predicted to induce collective flow effects\cite{greiner}.
In heavy-ion collisions at beam energies below 2 GeV/nucleon,
collective flow phenomena have been firmly
established by many experiments (e.g. \cite{dani88,gutbrod,westfall}).
In particular, the sideward deflection of matter in the reaction
plane was clearly revealed in the in-plane transverse
momentum analysis\cite{dani}. In this case, a typical
$S$ shape was seen for the average transverse momentum in the reaction
plane as a function of rapidity. It was also found that
the strength of the collective flow measured
in terms of the slope of the $S$ shaped distribution at mid-rapidity
or the total in-plane transverse momentum depends on the reaction 
dynamics, in-medium properties of hadron-hadron interactions 
and the equation of state\cite{sto,gale,bert,bauer92,bauer,pan,li93,zhang}. 
Furthermore, it was pointed out recently that a change in the strength of the 
collective flow could be a possible signature of quark-gluon
plasma formation in heavy ion collisions
at energies up to about 15 GeV/nucleon currently
available at the AGS\cite{qgsm,art1,dirk}. 
It is thus extremely interesting to test these predictions against
experimental data. However, it has been difficult to perform the 
standard in-plane transverse momentum analysis at the AGS due to the 
limited angular coverage and movability of existing spectrometers. Instead, 
azimuthal angle distributions of transverse 
energy measured with calorimeters without particle identification 
were analyzed\cite{e877}, and pronounced azimuthal anisotropies
indicating directed sideward flow were clearly identified.
Nevertheless, it is experimentally possible to study correlations between
transverse momenta of particles detected in the forward spectrometer 
and the flow direction determined by the calorimeters. 
It was thus first suggested in ref.\ \cite{e877a} that one can 
extract information about the transverse flow by studying the 
ratio $R(p_t)\equiv (dN^+/dp_t)/(dN^-/dp_t)$, where $N^+(N^-)$ is 
the number of particles in the reaction plane emitted in the 
same (opposite) direction of sideward flow, as a function of $p_t$ 
near the projectile rapidity.
Although this approach is similar in spirit to that for studying
the ``squeeze-out'' phenomenon at BEVALAC/SIS 
energies\cite{kaos,taps,li94,dani95}, it is still interesting to 
investigate whether this approach can reveal new 
features of the transverse flow and thus information about the 
equation of state of the superdense hadronic matter formed in 
relativistic heavy-ion collisions. In this paper, using both a simple, 
transversely moving thermal model and a more realistic, relativistic 
transport model (ART)\cite{art1} we show that the ratio $R(p_t)$ at 
high $p_t$ is very sensitive to the equation of state in the 
most violent stage of the reaction. 

Although particles are continuously emitted throughout the whole reaction 
process as indicated in our dynamical simulations, i.e, particles 
freeze-out at different temperatures, a simple thermal model with a 
single temperature is useful for a qualitative discussion of the more 
realistic dynamical calculations. We thus first discuss, using a transversely 
moving thermal model, the transverse momentum distributions 
$dN_{\pm}/p_tdp_t$ in the reaction plane for nucleons emitted 
in the same or opposite direction of the transverse flow. 
In particular, we discuss the limiting behaviour of the 
transverse momentum distributions and the ratio $R(p_t)$ at high $p_t$. 
  
Let us assume that all or a fraction of particles in a small rapidity 
bin around $y$ are in local thermal equilibrium at a temperature $T$,
and the center of mass of these particles are moving with velocity $\beta$ 
along the direction $+x$ in the reaction plane. 
From the standard transverse momentum analysis, we have
\begin{equation}
\beta=\frac{\sum_{i}(p_x)_i}{\sum_{i} E_i}
=\frac{\sum_{i}(p_x)_i}{\sum_{i}(m_t)_icosh(y)}
\leq \frac{\langle p_x\rangle }{m_ncosh(y)},
\end{equation}
where $\langle p_x\rangle$ is the average transverse momentum per nucleon 
in the reaction plane, and $m_n$ is the nucleon mass. 
For semicentral collisions of Au+Au at $P_{beam}/A=$ 10.8 GeV/c, 
$\langle p_x\rangle$ is about 0.1 GeV/c\cite{art1} 
at the projectile rapidity, $\beta$ thus has a value of about 0.05.
Using the Boltzmann distribution for the thermal source and boosting it with
the transverse velocity $\beta$, one obtains the following 
spectrum,
\begin{equation}\label{hh}
\frac{d^{3} N}{p_{t} dp_{t} d \phi dy}=C
\gamma (E-\beta p_t cos(\phi))
e^{-\gamma (E-\beta p_t cos(\phi))/T},
\end{equation}
where $C$ is the normalization constant and 
$\phi$ is the azimuthal angle with respect to the reaction plane. 
The transverse momentum spectra in a small rapidity bin around $y$ 
in the reaction plane for particles emitted in the same ($dN_{+}/p_t dp_t$) 
and opposite ($dN_{-}/p_t dp_t$) directions of the transverse flow 
are therefore given by
\begin{equation}
\frac{dN_{\pm}}{p_t dp_t}=C_{\pm}e^{-\gamma E/T}(\gamma E \mp T\alpha)
e^{\pm \alpha},
\end{equation}
where $\alpha\equiv \gamma \beta p_t/T$.
These distributions reduce to simple exponentials at high transverse momenta
\begin{equation}
(dN_{\pm}/p_tdp_t)_{\infty}\propto exp(-p_t/T^{\pm}_{eff}),
\end{equation} 
where the inverse slopes or effective 
temperatures $T^{\pm}_{eff}$ in the semilogarithmic plot of the spectra at 
high $p_t$ are obtained as
\begin{equation}\label{tem} 
\frac{1}{T^{\pm}_{eff}}=-lim_{p_t\rightarrow \infty}[\frac{d}{dp_t}
\ln(\frac{dN_{\pm}}{p_tdp_t})]
=\frac{\gamma}{T}(cosh(y)\mp \beta).
\end{equation}
In general, the inverse slope $1/T^{\pm}_{eff}$ reflects combined
effects of the temperature $T$ and the transverse flow velocity $\beta$.
For the special case of particles at midrapidity, $\beta$ is zero and 
the effective temperatures $T^{\pm}_{eff}$ equal to the local 
temperature $T$ of the thermal source. Otherwise, one expects
$T^{+}_{eff} > T^{-}_{eff}$. For high energy heavy-ion collisions
such as those at AGS energies, $\beta$ is much smaller than $cosh(y)$ 
around the projectile rapidity, and one again expects 
$T^{+}_{eff}\approx T^{-}_{eff}$, i.e. two approximately parallel spectra
$dN_{+}/p_t dp_{t}$ and $dN_{-}/p_tdp_t$ at high transverse momenta. 
This fact also indicates that it is difficult to extract the transverse 
flow velocity by studying the transverse momentum spectra alone. 

The degree of azimuthal asymmetry or the strength of transverse flow 
can be expressed in terms of the ratio
\begin{equation}
R(p_t)\equiv \frac{dN_{+}/dp_t}{dN_{-}/dp_t}=\frac{C_{+}}{C_{-}}\cdot
\frac{1-\beta\frac{p_t}{E}}{1+\beta\frac{p_t}{E}}\cdot 
e^{2p_t\frac{\gamma\beta}{T}}
\end{equation}    
To see how $R(p_t)$ might be sensitive to the flow velocity $\beta$ 
and/or the temperature $T$, we illustarte in Fig.\ 1 its variation
with $p_t$ for four sets of $\beta$ and $T$. It is seen that 
the ratio increases as a function of $p_t$ for any fixed values of $T$ 
and $\beta$. A slight increase in $\beta$ or decrease in $T$ results in 
a dramatic increase in the ratio $R(p_t)$. These results indicate that 
the $p_t$ dependence of the ratio $R(p_t)$ indeed carries interesting
information about the strength of transverse flow and the freeze-out 
temperature. It is, however, important to stress 
that in reality particles are emitted continuously at different freeze-out
temperatures during the whole reaction process. In particular, 
particles with high $p_t$ are mostly emitted from the most violent 
space-time regions where the local temperatures are high. 
Since the ratio $R(p_t)$ varies slowly with the transverse momentum for 
low $\beta/T$, one thus expects in dynamical model calculations 
a weaker $p_t$ dependence of the ratio $R(p_t)$ at high transverse momenta. 

We now turn to our study using the relativisitc transport 
model (ART 1.)\cite{art1}. This model was developed by including more 
baryon and meson resonances as well as their interactions in the BUU 
transport model (e.g. \cite{bert,li91a,li91b}). 
More specifically, we have included in ART 1.0 the following 
baryons: $N,~\Delta(1232),~N^{*}(1440),~N^{*}(1535),
~\Lambda,~\Sigma$, and mesons: $\pi,~\rho,~\omega,~\eta,~K$, 
as well as their explicit isospin degrees of freedom. Both elastic and 
inelastic collisions among most of these particles are included 
by using as inputs either the available experimental data or results
from one-boson-exchange and resonance models. An optional, self-consistent 
mean field for baryons is also included. We refer the reader to 
ref.\ \cite{art1} for more details of the model and its 
applications in studying various aspects of heavy-ion collisions at 
AGS energies.

In Fig. 2, we show the nucleon spectra, $dN_{\pm}/m_tdm_t$ versus $m_t-m$, 
for both central and peripheral collisions. To increase the statistics 
of our analysis particles with azimuthal angles smaller than $20^0$
with respect to the reaction plane are included. 
The chosen rapidity range $|y-2.95|\leq 0.35$ is very close to 
the projectile rapidity of 3.1. In this rapidity range particles 
with high transverse momenta must have suffered very violent
collisions and thus originate mostly from the very hot and dense 
participant region where large density gradients exist.
On the other hand, particles with low transverse momentum are mostly from the
cold spectators. In both central and peripheral collisions 
the spectra show typical exponential shapes for $m_t-m\geq 0.2$ GeV 
(or $p_t\geq 0.64$ GeV/c). Moreover, the spectra at high
$p_t$ for particles moving in the same and opposite directions 
of the transverse flow are approximately parallel to each other. 
These features are what we have expected from the schematic discussions
based on the transversely moving thermal model. 
We notice that the transverse momentum distribution in peripheral 
collisions is dominated by low transverse momentum particles from the 
spectators. From the spectra at high $p_t$, effective temperatures of 
about 150 MeV and 105 MeV are found for the central and peripheral collisions,
respectively. However, one needs to be cautious in interpreting 
these effective temperatures as particles are 
continuously freezed out at different temperatures during the whole 
reaction time. The effective temperatures determined from the 
spectra at high $p_t$ are therefore related through Eq.\ \ref{tem} to the 
most probable or average local temperatures in the most violent regions
of the reaction.

To study the strength of the transverse flow, we show in 
Fig.\ 3 the ratio $R(p_t)$ from our model calculations 
for the reaction of Au+Au at $p_{beam}/A=$10.8 GeV/c. As expected, 
the ratio increases gradually at low $p_t$ and reaches a limiting 
value at high $p_t$ in both central and peripheral collisions. Moreover, 
it is seen that the limiting value in peripheral collisions 
is much larger than that in central collisions. 
The reason for this behaviour is that even though the number of particles 
with high $p_t$ is less in peripheral collisions, these particles 
have a relatively higher value of $\beta/T$ due to a 
lower temperature. As the transverse momentum 
decreases $R(p_t)$ is increasingly more 
affected by particles from the cold spectators, 
and the ratio approaches one as $p_t$ goes to zero. 
This is most obvious for peripheral collisions where 
the isotropic Fermi motion of spectator nucleons
dominates and leads to $R(p_t)\approx 1$ for $p_t\leq$ 0.2 GeV/c.

Since $R(p_t)$ at high $p_t$ is a very sensitive measure of the
quantity $\beta/T$, and the local temperatures during the reaction 
process using the cascade and the soft equation of state are almost 
identical\cite{art1}, one expects that $R(p_t)$ depends strongly on 
the equation of state of the superdense hadronic matter formed 
in the most violent stage of the reaction. 
This is clearly seen also in Fig.\ 3 by comparing results obtained 
using the pure cascade and the soft equation of state for central collisions. 
Although there is no significant difference between 
the two values of $R(p_t)'s$ for $p_t\leq 0.2$ GeV/c,
they differ by about a factor of 1.5 at higher $p_t$. 
Moreover, the ratios are constants within a large range of high 
transverse momenta, and it can thus be relatively easily measured 
in experiments. 

In summary, we have investigated theoretically a new approach for measuring
the strength of transverse flow in terms of the number of 
particles in the reaction plane emitted in the same and opposite 
directions of the flow. The variation of the
strength of flow has been examined as a function of transverse momentum
using both a simple, transversely moving thermal model and a more
realistic, transport model. It is shown that the strength reaches 
a limiting value at high $p_t$, and this value is very sensitive to 
the equation of state. This new approach
provides new information about the collective flow that is
complementary to what one learns from the standard in-plane transverse 
momentum analysis but has the advantage of requiring particle 
identification only at a single rapidity. 

We would like to thank P. Danielewicz for many helpful discussions. 
The support of CMK by a Humboldt Research Award is also gratefully 
acknowledged, and he would like to thank
Ulrich Mosel of the University of Giessen for the warm hospitality.
This work was supported in part by NSF Grant No. PHY-9212209 
and PHY-9509266.

\newpage 
\section*{Figure Captions}
\begin{description}
 
\item{\bf Fig.\ 1}\ \ \
The strength of transverse flow as a function of transverse momentum 
predicted by a transversely moving thermal model. The temperature $T$ and
velocity $\beta$ are in units of GeV and c, respectively.

\item{\bf Fig.\ 2}\ \ \
Transverse mass distributions in the reaction plane for particles 
emitted in the same/opposite side of the transverse flow in reactions
of Au+Au at $p_{beam}/A=10.8$ GeV/c and at impact parameters less then
4 fm (upper window) and between 8 and 10 fm (lower window).

\item{\bf Fig.\ 3}\ \ \
Transverse momentum dependence of the strength of 
transverse flow in the reaction 
plane for the reactions described in Fig.\ 2.
Results using the soft equation of 
state and the pure cascade mode of ART are compared for 
central collisions.

\end{description}
 
\end{document}